%% file: main.tex
\documentclass[10pt, conference]{IEEEtran}
\IEEEoverridecommandlockouts
\usepackage{subfiles}
\usepackage{cite}
\usepackage{amsmath,amssymb,amsfonts}
\usepackage{algorithmic}
\usepackage{graphicx}
\usepackage{textcomp}
\usepackage{xcolor}
\usepackage{enumitem}
\usepackage{hyperref}
\usepackage{tabularx}
\usepackage{multirow}
\usepackage{booktabs}
\usepackage{tcolorbox}
\usepackage{etoolbox}
\usepackage{makecell}
\usepackage{mdframed}
\usepackage{xcolor, colortbl}
\usepackage{subcaption}
\hypersetup{nolinks=true}
\def\BibTeX{{\rm B\kern-.05em{\sc i\kern-.025em b}\kern-.08em
    T\kern-.1667em\lower.7ex\hbox{E}\kern-.125emX}}
\usepackage{listings}
\usepackage{xcolor}
\usepackage[T1]{fontenc}

\lstdefinelanguage{json}{
  basicstyle=\ttfamily\small,
  numbers=none,
  numberstyle=\tiny,
  stepnumber=1,
  numbersep=8pt,
  showstringspaces=false,
  breaklines=true,
  frame=single,
  morestring=[b]",
}

\begin{document}

\title{R2Code: A Self-Reflective LLM Framework for Requirements-to-Code Traceability%
\thanks{\textsuperscript{*}Corresponding author: Yishu Li.}
}

\author{
\IEEEauthorblockN{Yifei Wang}
\IEEEauthorblockA{\begin{minipage}[t]{0.30\textwidth}\centering
\textit{Dept of Computer Science} \\
\textit{City University of Hong Kong} \\
Hong Kong, China \\
ywang4748-c@my.cityu.edu.hk
\end{minipage}}
\and
\IEEEauthorblockN{Jacky Keung}
\IEEEauthorblockA{\begin{minipage}[t]{0.30\textwidth}\centering
\textit{Dept of Computer Science} \\
\textit{City University of Hong Kong} \\
Hong Kong, China \\
Jacky.Keung@cityu.edu.hk
\end{minipage}}
\and
\IEEEauthorblockN{Xiaoxue Ma}
\IEEEauthorblockA{\begin{minipage}[t]{0.30\textwidth}\centering
\textit{Dept of Electronic Engineering\\ and Computer Science} \\
\textit{Hong Kong Metropolitan University} \\
Hong Kong, China \\
kxma@hkmu.edu.hk
\end{minipage}}
\and
\IEEEauthorblockN{Zhenyu Mao}
\IEEEauthorblockA{\begin{minipage}[t]{0.35\textwidth}\centering
\textit{Dept of Computer Science} \\
\textit{City University of Hong Kong} \\
Hong Kong, China \\
zhenyumao2-c@my.cityu.edu.hk
\end{minipage}}
\and
\IEEEauthorblockN{Kehui Chen}
\IEEEauthorblockA{\begin{minipage}[t]{0.25\textwidth}\centering
\textit{Dept of Computer Science} \\
\textit{City University of Hong Kong} \\
Hong Kong, China \\
kehuichen2-c@my.cityu.edu.hk
\end{minipage}}
\and
\IEEEauthorblockN{\hspace{0.05\textwidth}Yishu Li\smash{\textsuperscript{*}}}
\IEEEauthorblockA{\hspace{0.05\textwidth}\begin{minipage}[t]{0.30\textwidth}\centering
\textit{Dept of Electronic Engineering\\ and Computer Science} \\
\textit{Hong Kong Metropolitan University} \\
Hong Kong, China \\
sliy@hkmu.edu.hk
\end{minipage}}
}

\maketitle

\begin{abstract}
Accurate requirement-to-code traceability is crucial for software maintenance. However, existing IR- and embedding-based methods are heavily dependent on lexical similarity, often yielding incomplete or inconsistent links across projects and languages and incurring high cost from long-context retrieval and prompting.
This paper presents R2Code, an LLM-based semantic traceability framework designed to improve trace link accuracy while reducing inference cost. R2Code integrates three components:
1) a decomposition-enhanced Bidirectional Alignment Network (BAN) that aligns four-layer requirement semantics with corresponding code structures to support cross-level semantic matching;
2) a Self-Reflective Consistency Verification (SRCV) module that conducts explanation-guided consistency checking to calibrate link reliability; and
3) a Dynamic Context-Adaptive Retrieval (DCAR) mechanism that adjusts retrieval granularity and filters contexts using semantic-overlap weighting for efficient context utilization.
Experiments on five public datasets spanning multiple domains and two programming languages demonstrate that R2Code consistently outperforms the strongest baselines, achieving an average F1 gain of 7.4\%, while reducing token consumption by up to 41.7\% through adaptive context control.
\end{abstract}


\input{1_Introduction}
\input{2_Background_RelatedWork}

\input{3_Methodology}

\input{4_Experiment}
\input{5_Results}
\input{6_Discussion}

\input{7_Conclusion}
\input{8_Ack}

\bibliographystyle{IEEEtran}
\bibliography{main}

\end{document}

%% file: 1_introduction.tex
\section{Introduction}
%% User story
Requirements-to-Code Traceability maintains explicit links among heterogeneous software artifacts such as requirements, design documents, source code, and test cases \cite{ramesh1997requirements}.
It links high-level, free-text documentation (e.g., requirements and design documents) to the source code elements that implement those requirements \cite{antoniol2025recovering}. 
Such links support change impact analysis and program comprehension, and they influence the effort required to maintain and evolve software systems \cite{ghabi2012code}. 
In practice, because traceability links are often missing or were not consistently recorded, an analyst may start from a natural-language requirements document and retrieve a ranked list of code elements that are likely to implement it.
However, maintaining traceability in evolving codebases remains challenging: manual tracing is labor-intensive and difficult to scale, and prior studies report that traceability maintenance may consume a substantial portion of overall development effort \cite{ramesh1998factors,tian2021impact}.

%% Existing methods 
Traditional information retrieval (IR) techniques, such as TF-IDF, BM25, VSM, LSI \cite{lyu2023systematic}, and WMD \cite{hey2021improving}, have long been used for automated trace link recovery. 
These methods compute similarity from lexical overlap or term statistics, enabling efficient retrieval but limiting their ability to capture deeper semantics \cite{guo2025natural}. 
Consequently, they often struggle with (i) surface-level matching without semantic understanding \cite{guo2025natural}, (ii) semantic and structural mismatches between high-level requirement descriptions and low-level code implementations \cite{merten2016information}, and (iii) missing explicit verification of semantic coherence in inferred links, which can lead to false positives.

Recent efforts explored large language models (LLMs) for requirements traceability, leveraging their reasoning and code understanding to go beyond keyword matching and infer requirement-to-code links \cite{north2024code}. 
Existing studies are preliminary, often relying on direct prompting or lightweight retrieval-augmented pipelines without traceability-specific decision mechanisms \cite{alor2025evaluating,zadenoori2025large}, while emerging work on engineering LLM-based systems also points to the need for more structured, protocol-driven designs\cite{mao2025multiagent}.
In real-world settings, traceability requires (i) structured alignment between multi-layer requirement semantics and code logic, and (ii) cost-effective context construction, as naively expanding context increases token cost and noise \cite{li2024simac}, which may degrade accuracy, efficiency, and robustness across projects and languages.

%% Problem
The core challenge of requirement-to-code traceability is to bridge the semantic and structural gap between human-oriented requirement intent and program-level operational behavior, robustly across domains and programming languages, while controlling the cost of context construction for LLM inference \cite{wang2025hgnnlink}. This motivates a framework that can explicitly align requirement and code semantics and selectively use the most relevant context for decision-making, enabling more reliable and efficient trace decisions in practice.

%% Approach
To address these challenges, we present R2Code, an LLM-based framework for requirements-to-code traceability that targets three core objectives: effectiveness, robustness, and efficiency.
R2Code bridges requirement--code mismatches via explicit cross-level semantic alignment, improves link reliability through self-reflective consistency checking, and reduces inference cost by constructing compact evidence with context-adaptive retrieval.
We evaluate R2Code on five public datasets spanning multiple domains and two programming languages, demonstrating consistent F1 improvements over strong IR, dense retrieval, and RAG-based baselines while substantially reducing token consumption.

The main contributions of this paper include:
%% Contribution
\begin{itemize}
    \item \textbf{Framework.} We propose an end-to-end evidence-to-decision framework for requirement-to-code traceability with three key designs: bidirectional semantic alignment, explanation-guided confidence calibration, and cost-aware adaptive evidence construction.

    \item \textbf{Empirical study.} We conduct an extensive evaluation on five public datasets across multiple domains and two programming languages, comparing against strong IR, dense retrieval, and RAG-based baselines, and demonstrate consistent F1 gains with substantial reductions in token consumption.
\end{itemize}

%% file: 2_Background_RelatedWork.tex
\section{Background and RelatedWork}
\subsection{Requirements-to-Code Traceability: Task Overview}
Requirements-to-code traceability aims to establish and maintain explicit links between natural-language requirement statements and the source code entities (e.g., files, classes, or methods) that implement them \cite{cleland2014software, ramesh2002toward}.
Such requirement--implementation links provide essential support for software maintenance and evolution tasks, including change impact analysis, program comprehension \cite{niu2016gray}. 
In practice, however, trace links are frequently missing, outdated, or inconsistent in evolving codebases \cite{rasiman2022effective}, which motivates automated techniques for trace link recovery \cite{aung2020literature}.

Formally, given a set of requirements $R = \{ r_1, \ldots, r_m \}$ and a set of code entities $C = \{ c_1, \ldots, c_n \}$, the goal is to recover a trace link set $L \subseteq R \times C$ that approximates a ground-truth set $L^{*}$ \cite{zhang2016empirical}. 
Prior work formulates the task in two modes: \emph{ranking}, which returns a ranked list of candidate code entities for each requirement, and \emph{classification} \cite{wang2022systematic}, which predicts whether a requirement--code pair constitutes a valid link under a similarity or confidence threshold \cite{hey2024requirements}. 
Accordingly, evaluation reports Precision, Recall, and F1 for link prediction \cite{hayes2003improving}, and may also report ranking-based metrics (e.g., MAP/MRR) when retrieval quality is emphasized \cite{zhang2023ealink}.

\subsection{Classical IR-based Methods}
Classical information retrieval (IR) approaches are widely used as baselines for automated trace link recovery \cite{borg2014recovering}. They typically follow a common pipeline: (i) represent each requirement and each code entity as a textual artifact (e.g., derived from requirement descriptions and code identifiers/comments); (ii) compute a relevance score $s(r,c)$ from term statistics or vector-space similarity; and (iii) rank candidate code entities $c \in C$ for a given requirement $r$, or convert scores into binary links via thresholding \cite{hayes2003improving}.

Representative IR baselines in traceability include TF-IDF and Vector Space Model (VSM) variants, as well as BM25-style ranking functions \cite{robertson2009probabilistic}, which estimate relevance from term-frequency statistics (e.g., TF, IDF, and length normalization) \cite{robertson2009probabilistic}. 
Latent Semantic Indexing (LSI) projects documents into a low-dimensional latent space via matrix factorization, capturing co-occurrence patterns beyond surface term overlap \cite{deerwester1990indexing}. 
Word Mover's Distance (WMD) and related distance-based measures compute document similarity using distances in a word-embedding space, enabling softer semantic matching than exact term overlap \cite{Kusner2015WMD}.

Early traceability research instantiated the IR paradigm with a range of relevance models. 
Marcus and Maletic leveraged LSI to recover documentation-to-code links by ranking candidate code artifacts in a latent semantic space \cite{marcus2003recovering}. 
Sundaram et al. evaluated VSM with alternative term-weighting schemes (e.g., TF-IDF/Okapi) alongside LSI across multiple datasets \cite{sundaram2010assessing}.
Oliveto et al. analyzed the overlap and equivalence of candidate links produced by IR techniques\cite{oliveto2010equivalence}.
And Hey et al. explored WMD-based semantic distance for traceability link identification \cite{hey2021improving}.

\subsection{Neural Retrieval and LLM-based Methods}
Beyond classical IR, a large body of work formulates trace link recovery with neural retrieval models that learn semantic representations of requirements and code artifacts \cite{guo2017semantically,lin2021traceability}. A common approach is the bi-encoder (dual-encoder) architecture, which encodes a requirement $r$ and a code entity $c$ independently into dense vectors and computes similarity (e.g., dot product or cosine similarity) for ranking \cite{reimers2019sentence}. This design enables efficient nearest-neighbor search over large candidate sets and is often used as a first-stage retriever. In addition, cross-encoder or interaction-based models are sometimes adopted as rerankers: they jointly encode $(r,c)$ and predict a relevance score, typically trading off higher inference cost for more fine-grained matching \cite{leonhardt2024efficient}.

Neural retrieval for traceability often leverages pretrained Transformer encoders to represent requirement text and textualized code artifacts \cite{lin2021traceability}. 
Requirements are encoded as sentences or short documents, while code entities are represented using identifiers, comments, and related textual signals; code-pretrained models can further improve the capture of programming-language semantics \cite{feng2020codebert,guo2022unixcoder}. 
These encoders can be fine-tuned with supervision from known trace links to improve artifact-level similarity estimation \cite{tian2023cross}. 
Similar to IR pipelines, neural approaches support both top-$k$ ranked retrieval and threshold-based link prediction under standard evaluation settings \cite{guo2017semantically,lin2021traceability}.

More recently, LLM-based methods have been explored for traceability by leveraging their reasoning capabilities to model requirement-to-code relationships and generate natural-language rationales for predicted links~\cite{fuchss2025lissa,wang2025chart2code}.
A common pattern is retrieval-assisted inference: a retriever (IR or neural) first selects a small set of candidate code entities, and the LLM then performs link assessment, sometimes with explanation generation or verification-style prompting, to output the final decision~\cite{hey2025requirements,ali2024establishing}. In this way, LLMs serve as a semantic decision module on top of retrieved evidence, complementing retrieval-centric approaches \cite{li2024llm}.

\subsection{Research Gap}
Prior work on requirements-to-code trace link recovery offers strong baselines, yet key gaps remain when jointly optimizing accuracy, robustness, and efficiency \cite{alor2025evaluating,mao2025re}. 
\textbf{(i) Effectiveness: Cross-level mismatch.} 
Classical IR and neural retrieval approaches largely operationalize traceability as text similarity scoring (lexical matching, latent semantic projection, or dense embeddings), which supports ranking but does not explicitly capture the correspondence between high-level requirement intent and distributed implementation logic in code. 
\textbf{(ii) Robustness: Uncalibrated confidence.} 
Retrieval-centric pipelines typically output similarity scores without an explicit mechanism to verify semantic coherence, limiting interpretability and allowing spurious matches to propagate.
\textbf{(iii) Efficiency: Evidence cost.} 
Recent LLM-based studies often rely on direct prompting or lightweight retrieval-augmented pipelines whose outcomes are sensitive to evidence selection and presentation \cite{north2024code,hey2025requirements}. Static or overly broad context increases inference cost and introduces irrelevant information, while overly narrow context may omit crucial implementation cues \cite{li2024simac}. 
These limitations motivate traceability designs that integrate structured alignment, calibrated confidence, and cost-aware evidence construction.

%% file: 3_Methodology.tex
\section{Methodology}
\subsection{Overview}
Fig.~\ref{fig:r2code} summarizes the R2Code workflow. Given a requirement $r$ and a codebase $C$, R2Code ranks candidate code entities and outputs trace links with calibrated confidence scores. 
R2Code adopts a two-stage scoring process: it derives structured semantic representations of requirements and code and computes an alignment score $s_{\mathrm{BAN}}(r,c)$ via bidirectional semantic alignment (Sec.~III-B), then refines it into a calibrated confidence $s_{\mathrm{final}}(r,c)$ through self-verification (Sec.~III-C). R2Code also generates an explanation $E(r,c)$ to support human validation.
In parallel, a context-adaptive retrieval mechanism (Sec.~III-D) selects compact and relevant evidence for LLM inference, reducing redundant context while preserving salient signals. R2Code comprises three components:
\begin{figure*}[hbtp]
    \centering
    \includegraphics[width=\textwidth]{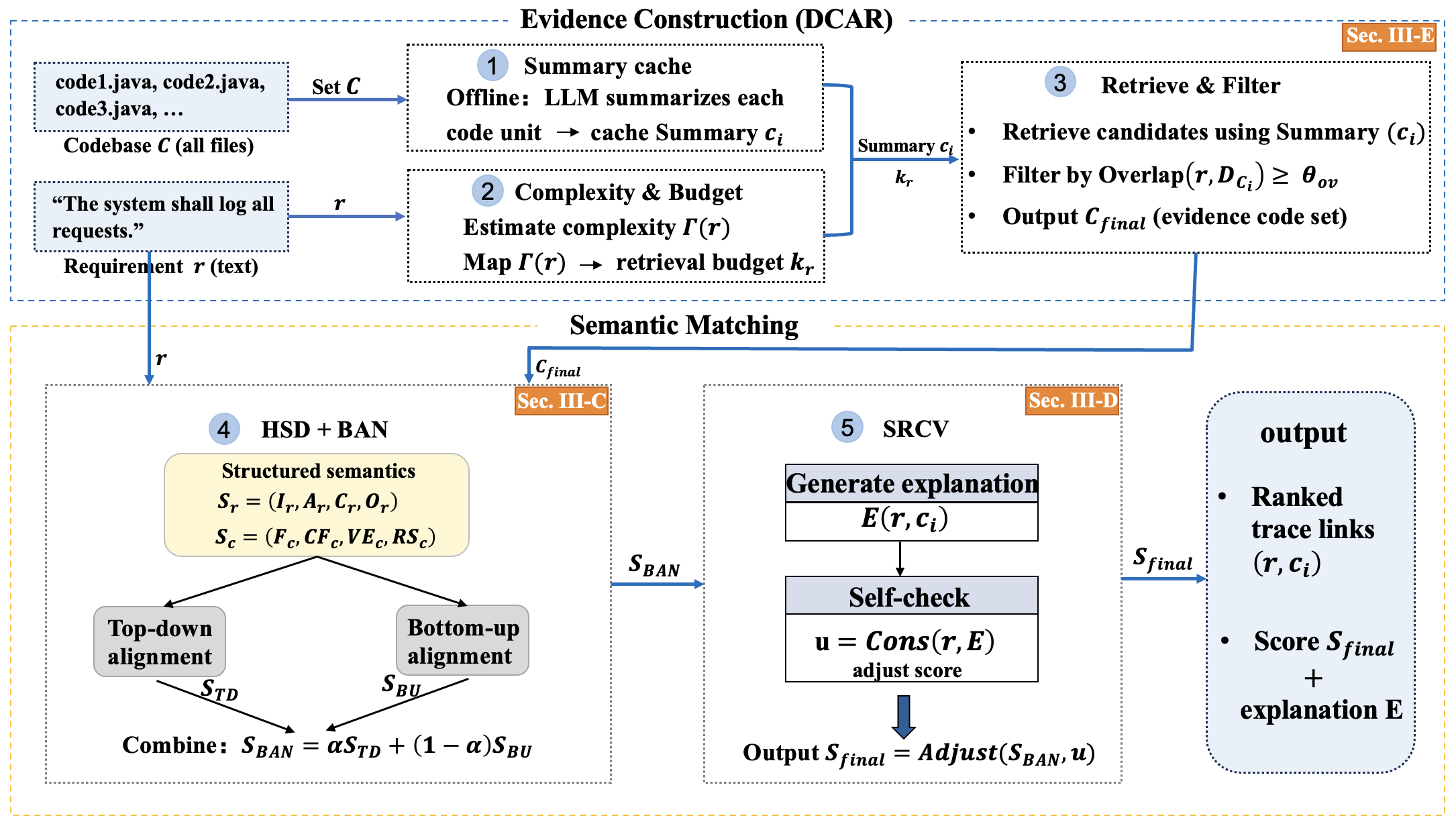}
    \caption{R2Code Traceability Workflow}
    \label{fig:r2code}
\end{figure*}

\textbf{Decomposition-Enhanced BAN (HSD+BAN).}
R2Code first applies HSD to decompose each requirement and summarize each candidate code entity into comparable multi-layer semantic representations, and then uses BAN to perform top-down and bottom-up alignment to compute an initial matching score $s_{\mathrm{BAN}}(r,c)$.

\textbf{Self-Reflective Consistency Verification (SRCV).}
To improve reliability, SRCV generates an explanation for a predicted link and verifies its consistency with the requirement. The verification result calibrates the initial score, yielding the final confidence $s_{\mathrm{final}}(r,c)$ for link decision-making.

\textbf{Dynamic Context-Adaptive Retrieval (DCAR).}
DCAR governs evidence construction for LLM inference by adapting retrieval granularity and filtering context based on semantic overlap. It aims to provide compact yet informative context to downstream reasoning, improving efficiency without sacrificing link quality.

\subsection{Problem Formulation}
This work considers the requirements-to-code traceability problem between a set of natural-language requirements and a source code corpus. Let
\[
R=\{r_1,\ldots,r_m\}
\]
denote the requirement set, and
\[
C=\{c_1,\ldots,c_n\}
\]
denote the set of code entities, where each $c\in C$ can be a file, class, or method. The goal is to recover a ground-truth link set
\[
L^\ast \subseteq R \times C,
\]
where $(r,c)\in L^\ast$ indicates that $c$ (fully or partially) implements $r$. Given this definition, a traceability approach defines a scoring function
\[
s: R \times C \rightarrow [0,1]
\]
and produces predicted links by thresholding:
\[
\hat{L}=\{(r,c)\mid r\in R,\ c\in C,\ s(r,c)\ge \theta\},
\]
where $\theta\in[0,1]$ is a decision threshold.

We model requirements-to-code traceability as a parameterized evidence-to-decision mapping that assigns each requirement--code pair a matching score and an auxiliary explanation for validation.
For each pair $(r,c)$, the framework outputs an alignment score $s_{\mathrm{BAN}}(r,c)$, a calibrated confidence $s_{\mathrm{final}}(r,c)$, and an explanation $E(r,c)$ describing the rationale for accepting or rejecting the link. Overall, R2Code can be viewed as a parameterized mapping
\[
\begin{aligned}
f_{\Theta}:(R,C)\rightarrow 
\{&(r,c,s_{\mathrm{BAN}}(r,c),s_{\mathrm{final}}(r,c),E(r,c)) \\
&\mid r\in R,\ c\in C\},
\end{aligned}
\]
where $\Theta$ denotes the framework configuration. $s_{\mathrm{final}}(r,c)$ is designed to approximate the unknown traceability relation encoded in $L^\ast$, while $E(r,c)$ provides interpretable support for downstream verification.

\subsection{HSD+BAN: Decomposition and Alignment}
R2Code begins by converting both requirements and code entities into structured semantic representations, so that matching is performed over aligned semantic dimensions rather than raw text alone. Specifically, the framework applies Hierarchical Semantic Decomposition (HSD) to obtain a four-layer requirement representation and a four-layer code representation, and then uses a Bidirectional Alignment Network (BAN) to compute an initial alignment score $s_{\text{BAN}}(r,c)$ via complementary top-down and bottom-up reasoning.

\paragraph{Requirement semantic decomposition}
For each requirement $r$, HSD decomposes its semantics into four dimensions: intent, actions, conditions, and outputs
\begin{equation}
S_r = (I_r, A_r, C_r, O_r).
\end{equation}
Here, $I_r$ denotes the intent, $A_r=\{a_i\}$ the required actions, $C_r=\{c_i\}$ the conditions and constraints, and $O_r=\{o_i\}$ the expected outputs.

Given requirement text $T_r$, we prompt the LLM to produce $S_r$ in a structured JSON format with the corresponding fields (\texttt{intent}, \texttt{actions}, \texttt{conditions}, \texttt{outputs}). We cap each list at 5 concise phrases and set $temperature=0$ for deterministic parsing.

Example (simplified): For \textit{``The system shall log all requests''}, the output is:
\begin{lstlisting}
{
  "intent": ["Maintain an audit trail for incoming requests"],
  "actions": ["record request metadata", "persist log entry"],
  "conditions": ["for every received request"],
  "outputs": ["a stored log entry that can be queried later"]
}
\end{lstlisting}

\paragraph{Code semantic decomposition}
For each code entity $c$ (with textual view $T_c$), HSD constructs a corresponding four-layer representation
\begin{equation}
S_c = (F_c, CF_c, VE_c, RS_c),
\end{equation}
where $F_c$ captures the function's intent, $CF_c=\{cf_i\}$ represents control-flow structures (e.g., branches and loops), $VE_c=\{ve_i\}$ denotes variable effects and state changes, and $RS_c=\{rs_i\}$ encodes return values or final states.
Table~\ref{tab:layer-mapping} summarizes the four-layer correspondence between requirement and code semantics used in BAN.
\begin{table}[t]
\caption{Requirement–Code Layer Mapping}
\label{tab:layer-mapping}
\centering
\begin{tabular}{l l l}
\hline
\textbf{Requirement layer} & \textbf{Code layer} & \textbf{What it captures} \\
\hline
Intent $I_r$ & Function intent $F_c$ & goal / responsibility \\
Actions $A_r$ & Control flow $CF_c$ & operations / logic steps \\
Conditions $C_r$ & Variable effects $VE_c$ & constraints / state changes \\
Outputs $O_r$ & Return states $RS_c$ & expected results / outputs \\
\hline
\end{tabular}
\end{table}

\paragraph{Bidirectional Alignment Network (BAN)}
BAN estimates the semantic correspondence between $S_r$ and $S_c$ through two complementary directions. Top-down alignment focuses on requirement satisfaction by assessing whether the code covers the requirement's intent and expectations:
\begin{equation}
\begin{aligned}
s_{\mathrm{TD}}=\frac{1}{4}\bigl(
&\mathrm{sim}(I_r,F_c)+\mathrm{sim}(A_r,CF_c) \\
&+\mathrm{sim}(C_r,VE_c)+\mathrm{sim}(O_r,RS_c)
\bigr).
\end{aligned}
\end{equation}
Bottom-up alignment evaluates implementation fidelity by verifying that the code’s logic aligns with the requirement:
\begin{equation}
\begin{aligned}
s_{\mathrm{BU}}=\frac{1}{4}\bigl(
&\mathrm{sim}(F_c,I_r)+\mathrm{sim}(CF_c,A_r) \\
&+\mathrm{sim}(VE_c,C_r)+\mathrm{sim}(RS_c,O_r)
\bigr).
\end{aligned}
\end{equation}

$\mathrm{sim}(x,y)\in[0,1]$ is operationalized as a normalized semantic alignment score produced by the backbone LLM under a fixed prompt and a constrained output schema.
For each semantic layer pair $(x,y)$ (e.g., $(I_r, F_c)$, $(A_r, CF_c)$), the LLM performs structured joint reasoning to assess whether $y$ semantically supports $x$, and returns a normalized confidence score.
To ensure consistency and interpretability of the alignment process, the output is constrained to a structured JSON format.
Concretely, the LLM must return JSON in the following schema:
\begin{lstlisting}
{
  "score": <float>,
  "coverage": <"covered" | "partially_covered" | "not_covered">,
  "evidence": [<string>, ...]
}

\end{lstlisting}
The numeric field ``score'' is directly used as $\mathrm{sim}(x,y)$.
The remaining fields are optional and used only for interpretability and error analysis.

The final BAN score combines the two directions:
\begin{equation}
s_{\text{BAN}}(r,c) = \alpha\, s_{\text{TD}} + (1-\alpha)\, s_{\text{BU}},
\end{equation}
where $\alpha\in[0,1]$ balances requirement satisfaction (top-down verification) and implementation fidelity (bottom-up validation). The resulting $s_{\text{BAN}}(r,c)$ serves as the initial alignment score and is passed to the next stage for confidence calibration via self-verification (Section~III-D).

\subsection{SRCV: Self-Reflective Consistency Verification}
BAN provides an initial semantic alignment score $s_{\mathrm{BAN}}(r,c)$ for each requirement--code pair. However, LLM-based alignment may still exhibit over-confident links when the model forms plausible but unsupported associations. SRCV introduces a lightweight self-verification step that uses the model's own explanation as evidence and calibrates the confidence of $s_{\mathrm{BAN}}$ accordingly, producing the final score $s_{\mathrm{final}}$.

\paragraph{Explanation Generation}
Given $(r,c)$ and $s_{\mathrm{BAN}}(r,c)$, SRCV first asks the LLM to generate an explicit justification:
\begin{equation}
E(r,c)=\mathrm{LLM}_{\mathrm{gen}}(r,c,s_{\mathrm{BAN}}),
\end{equation}
where $E(r,c)$ summarizes the claimed semantic correspondence (e.g., which requirement intents are supported by which code behaviors). This step does not aim at making the output ``look reasonable''; rather, it externalizes the rationale into a verifiable object for subsequent checking.

\paragraph{Consistency Reflection}
SRCV then evaluates whether the explanation is consistent with the original requirement text by scoring:
\begin{equation}
\mathrm{Cons}(r,E)=\mathrm{LLM}_{\mathrm{judge}}(r,E)\in[0,1],
\end{equation}

To compute $\mathrm{Cons}(r,E)$, the backbone LLM is prompted to assess whether the generated explanation $E(r,c)$ is consistent with the original requirement text $T_r$.
The judge must return JSON in the following schema:
\begin{lstlisting}
{
  "consistent": <"yes" | "no">,
  "score": <float>,
  "reasons": [<string>, ...]
}
\end{lstlisting}
We define $\mathrm{Cons}(r,E)\in[0,1]$ as the numeric field ``score'', which is used for confidence calibration in Eq.~(9).

As a lightweight auxiliary signal, SRCV can additionally compute a term-level overlap between the requirement and the explanation:
\begin{equation}
\mathrm{Overlap}(r,E)=
\frac{|\mathrm{Terms}(r)\cap \mathrm{Terms}(E)|}
{|\mathrm{Terms}(r)\cup \mathrm{Terms}(E)|}.
\end{equation}
In our framework, $\mathrm{Overlap}(r,E)$ is treated only as supportive evidence rather than a primary decision factor.

\paragraph{Score Adjustment}
Finally, SRCV calibrates the BAN score using the consistency signal:
\begin{equation}
s_{\mathrm{final}}=\mathrm{Adjust}\bigl(s_{\mathrm{BAN}},\mathrm{Cons}(r,E)\bigr).
\end{equation}
High consistency slightly strengthens confidence, moderate consistency applies a mild penalty, and low consistency triggers a stronger penalty. Importantly, SRCV is not a separate predictive model; it is a confidence calibration layer that reduces false positives caused by hallucinated connections while preserving links supported by coherent reasoning.

\subsection{DCAR: Dynamic Context-Adaptive Retrieval}
DCAR controls evidence construction for LLM inference by adapting retrieval granularity and filtering noisy context on a per-requirement basis. Rather than using a fixed retrieval window for all requirements, DCAR (i) builds compact, reusable summaries for code units, (ii) allocates a requirement-specific retrieval budget, and (iii) retains only the most relevant items as input evidence for downstream reasoning.

\paragraph{Code Summarization (Cached)}
To avoid repeatedly sending full code into the LLM, DCAR generates a compact summary for each code unit $c$ (e.g., file/class/function) with text $T_c$:
\begin{equation}
\mathrm{Summary}(c)=(F_c, D_c, CC_c, \gamma_c),
\end{equation}
where $F_c$ is the function/name signature, $D_c$ is a short semantic description of the core behavior, $CC_c$ is the set of invoked/dependent functions (call links), and $\gamma_c\in[0,1]$ is an estimated code complexity. Summaries are cached and reused across requirements.

\paragraph{Requirement Complexity}
Given a requirement $r$ with text $T_r$, DCAR computes a semantic complexity score:
\begin{equation}
\Gamma(r)=w_1\phi_{\mathrm{len}}(T_r)+w_2\phi_{\mathrm{act}}(T_r)+w_3\phi_{\mathrm{cond}}(T_r)+w_4\phi_{\mathrm{conn}}(T_r),
\end{equation}
where $\phi_{\mathrm{len}},\phi_{\mathrm{act}},\phi_{\mathrm{cond}},\phi_{\mathrm{conn}}$ capture requirement length, action density, conditional indicators, and logical connectors, respectively.

\paragraph{Adaptive Retrieval Window}
Based on $\Gamma(r)$, DCAR sets a retrieval budget $k_r$ using a minimal piecewise rule:
\begin{equation}
k_r=
\begin{cases}
k_0,  & \Gamma(r)<\tau_1,\\
2k_0, & \tau_1\le \Gamma(r)<\tau_2,\\
3k_0, & \Gamma(r)\ge \tau_2,
\end{cases}
\end{equation}
where $k_0$ is the base retrieval size and $\tau_1,\tau_2$ are complexity thresholds. Simpler requirements retrieve a smaller set of function-level summaries, while more complex requirements may expand retrieval to include call-chain-related context via $CC_c$, so that distributed implementations can be covered.

\paragraph{Semantic Filtering}
After retrieval, DCAR further filters candidate contexts using a lightweight semantic-overlap signal computed between the requirement and each candidate’s summary text:
\begin{equation}
\mathrm{Overlap}(r,D_{c_i})=
\frac{|\mathrm{Terms}(r)\cap \mathrm{Terms}(D_{c_i})|}
     {|\mathrm{Terms}(r)\cup \mathrm{Terms}(D_{c_i})|}.
\end{equation}
Only items above a threshold are retained, with a final cap on context size:
\begin{equation}
C_{\mathrm{final}}=\{c_i \mid \mathrm{Overlap}(r,D_{c_i})\ge \theta_{\mathrm{ov}}\},\quad
|C_{\mathrm{final}}|\le k_{\max}.
\end{equation}
The resulting $C_{\mathrm{final}}$ constitutes the compact evidence passed to the downstream LLM decision stage.

%% file: 4_Experiment.tex
\section{Experiment}
\subsection{Research Questions}
This study evaluates R2Code through three research questions that reflect the core objectives of the framework: effectiveness, robustness, and efficiency.
\begin{itemize}
\item \textbf{RQ1 (Effectiveness):} How effectively does R2Code improve traceability accuracy compared with baseline methods, and how do its key components contribute to this improvement?
\item \textbf{RQ2 (Robustness):} Does R2Code maintain stable performance across different project domains and programming languages?
\item \textbf{RQ3 (Efficiency):} What efficiency gains does R2Code provide in terms of token usage and inference cost?
\end{itemize}

To address RQ1, we evaluate R2Code against traditional IR baselines (BM25, TF-IDF, VSM, LSI, WMD), a dense retriever, and a standard RAG+LLM pipeline on five benchmark datasets \cite{borg2014recovering, hey2025requirements}. We report precision, recall, and F1-score, and include ranking metrics ( MRR, and precision/recall@k) to assess retrieval quality\cite{hayes2003improving,zhang2023ealink}. We further examine the contributions of key components by comparing R2Code with its ablated variants.

To address RQ2, we evaluate R2Code on five public datasets spanning different application domains and two programming languages (Java and C\#), including four CoEST Java datasets~\cite{hey2021improving} and the RETRO.NET dataset for C\#~\cite{hayes2018requirements}.

To address RQ3, we record token consumption, inference latency, and estimated cost under different retrieval strategies. We compare fixed-window RAG~\cite{lewis2020retrieval} with the proposed dynamic context-adaptive retrieval (DCAR) to quantify the cost–accuracy trade-off.

\subsection{Settings}
\subsubsection{Dataset}
\begin{table}[htbp]
\centering
    \caption{Dataset Statistics} 
    \begin{tabular}{c|c|c|c|c|c}
    \toprule
        Dataset & Lang. & Description & Req. & Code & Gold Links\\
    \midrule
        iTrust & Java & Medical & 335 & 1,818 & 1,156 \\
        eTour & Java & Tourism & 316 & 1,337 & 1,162 \\
        SMOS & Java & Sensor & 154 & 1,220 & 552 \\
        eANCI & Java & Admin. & 66 & 316 & 74 \\
        RETRO.NET & C\# & Library & 66 & 118 & 301 \\
    \bottomrule
    \end{tabular}
    \label{tab:1}
\end{table}
To ensure a robust and reproducible evaluation of R2Code across diverse project domains and programming languages, we conduct experiments on five public requirement-to-code traceability datasets. 
As summarized in Table~\ref{tab:1}, these datasets cover medical systems, tourism platforms, sensor networks, and public administration, and span two programming languages (Java and C\#). Each dataset contains natural-language requirements, the corresponding source code files, and manually curated ground-truth traceability links.

\subsubsection{Model}
R2Code is implemented with DeepSeek-V3.1-Terminus as the backbone LLM for hierarchical decomposition, bidirectional alignment, and consistency verification. 
We adopt DeepSeek-V3.1-Terminus because DeepSeek's JSON Output allows us to enforce schema-constrained structured decomposition and normalized alignment scoring, enabling reliable parsing and downstream computation \cite{deepseek_json_output}.
For the baseline RAG pipeline, dense retrieval is implemented using sentence-transformers/all-mpnet-base-v2. Classical IR baselines include BM25, TF-IDF, VSM, LSI, and WMD, each configured with standard parameter settings.

\subsubsection{Environment and Parameters}
All experiments are conducted on an NVIDIA A100 server using Python 3.9 and PyTorch 2.0, with the random seed fixed to 42. 
The backbone LLM uses temperature 0.0, top-p 0.95, and max tokens 2048. 
IR baselines follow commonly used configurations in standard toolkits.
Within R2Code, BAN and SRCV use a fixed configuration throughout all experiments. For DCAR, we set $k_{\text{base}} = 5$ and cap dynamic expansion with $k_{\max} = 10$ to bound retrieval context and cost.

\subsubsection{Evaluation Metrics}
We evaluate traceability performance from three perspectives: (i) \emph{classification quality}, reported with Precision, Recall, and F1-score on predicted requirement--code links; (ii) \emph{ranking quality}, assessed by MRR along with Precision@k and Recall@k $(k \in \{5,10\})$ to reflect the quality of top-ranked candidates;
and (iii) \emph{efficiency}, measured by total input/output token consumption, the number of LLM requests, and end-to-end runtime, with inference cost estimated based on per-token pricing for input and output tokens. All results are averaged over requirements, and we report standard deviations where appropriate.

%% file: 5_Results.tex
\section{Results}
In this section, we present experimental results organized by the research questions in Section IV-A. We first report the accuracy of R2Code on the primary benchmark setting (iTrust) and analyze the contribution of key components via ablation (\textbf{RQ1}). We then examine robustness by evaluating the same configuration across multiple datasets spanning different application domains and programming languages (\textbf{RQ2}). Finally, we analyze efficiency in terms of token consumption, runtime, and estimated cost, and quantify the cost--accuracy trade-off under different retrieval strategies (\textbf{RQ3}).

\subsection{Answer to RQ1: Effectiveness}
\begin{table}[htbp]
\centering
\caption{Overall Performance on the iTrust Dataset}
\begin{tabular}{l|c|c|c|c|c|c}
\toprule
Method & F1 & Recall & Prec. & MRR & P@10 & R@10 \\
\midrule
BM25    & 0.6158 & 0.6084 & 0.6234 & 0.7234 & 0.5845 & 0.5234 \\
TF-IDF  & 0.6068 & 0.6014 & 0.6123 & 0.7089 & 0.5712 & 0.5123 \\
VSM     & 0.5928 & 0.5876 & 0.5981 & 0.6924 & 0.5568 & 0.4987 \\
LSI     & 0.5684 & 0.5623 & 0.5746 & 0.6689 & 0.5321 & 0.4715 \\
WMD     & 0.5910 & 0.5944 & 0.5876 & 0.6945 & 0.5523 & 0.3894 \\
RAG-LLM & 0.6949 & 0.6783 & 0.7123 & 0.8123 & 0.6823 & 0.6123 \\
R2Code  & \textbf{0.7296} & \textbf{0.7050} & \textbf{0.7560} & \textbf{0.8550} & \textbf{0.7720} & \textbf{0.7060} \\
\bottomrule
\end{tabular}
\label{tab:rq1_overall}
\end{table}

To answer RQ1, we compare R2Code with classical IR baselines and a standard RAG+LLM pipeline on iTrust.
As shown in Table~\ref{tab:rq1_overall}, R2Code achieves the best overall performance, reaching an F1-score of 0.7296 with 0.7560 precision and 0.7050 recall. Compared with the strongest baseline RAG+LLM (F1=0.6949), R2Code improves F1 by +0.0341, while also achieving higher ranking quality (MRR: 0.8550 vs. 0.8123). Under practical top-$k$ cutoffs, R2Code yields the highest P@10 (0.7720) and R@10 (0.7060), indicating more correct links can be surfaced earlier in the ranked list.

\begin{table}[htbp]
\centering
\caption{Ablation Study on the iTrust Dataset}
\begin{tabular}{l|c c c|c|c|c}
\toprule
Ablation & BAN & SRCV & DCAR & F1 & Prec. & Recall \\
\midrule
Full      & $\checkmark$ & $\checkmark$ & $\checkmark$ & \textbf{0.7296} & \textbf{0.7560} & \textbf{0.7050} \\
-- BAN    & $\times$     & $\checkmark$ & $\checkmark$ & 0.6949 & 0.7123 & 0.6783 \\
-- SRCV   & $\checkmark$ & $\times$     & $\checkmark$ & 0.6864 & 0.7873 & 0.6084 \\
-- DCAR   & $\checkmark$ & $\checkmark$ & $\times$     & 0.7112 & 0.8117 & 0.6329 \\
\midrule
BAN only  & $\checkmark$ & $\times$     & $\times$     & 0.6859 & 0.7600 & 0.6250 \\
SRCV only & $\times$     & $\checkmark$ & $\times$     & 0.6854 & 0.7234 & 0.6512 \\
DCAR only & $\times$     & $\times$     & $\checkmark$ & 0.6930 & 0.8122 & 0.6040 \\
\bottomrule
\end{tabular}
\captionsetup{justification=raggedright,singlelinecheck=false}
\caption*{\footnotesize \textbf{Note.} $\checkmark$ indicates the component is enabled; $\times$ indicates it is removed.}
\label{tab:ablation}
\end{table}
Table~\ref{tab:ablation} reports the ablation results of R2Code on iTrust. Removing BAN decreases the F1-score from 0.7296 to 0.6949. Disabling SRCV results in a larger drop (F1=0.6864), accompanied by a reduction in recall to 0.6084. When DCAR is removed, the F1-score remains close to the full model (0.7112), while recall drops from 0.7050 to 0.6329. Overall, the full configuration achieves the best balance among precision, recall, and F1 on iTrust.

\begin{figure}[htbp]
    \centering
    \includegraphics[width=\linewidth]{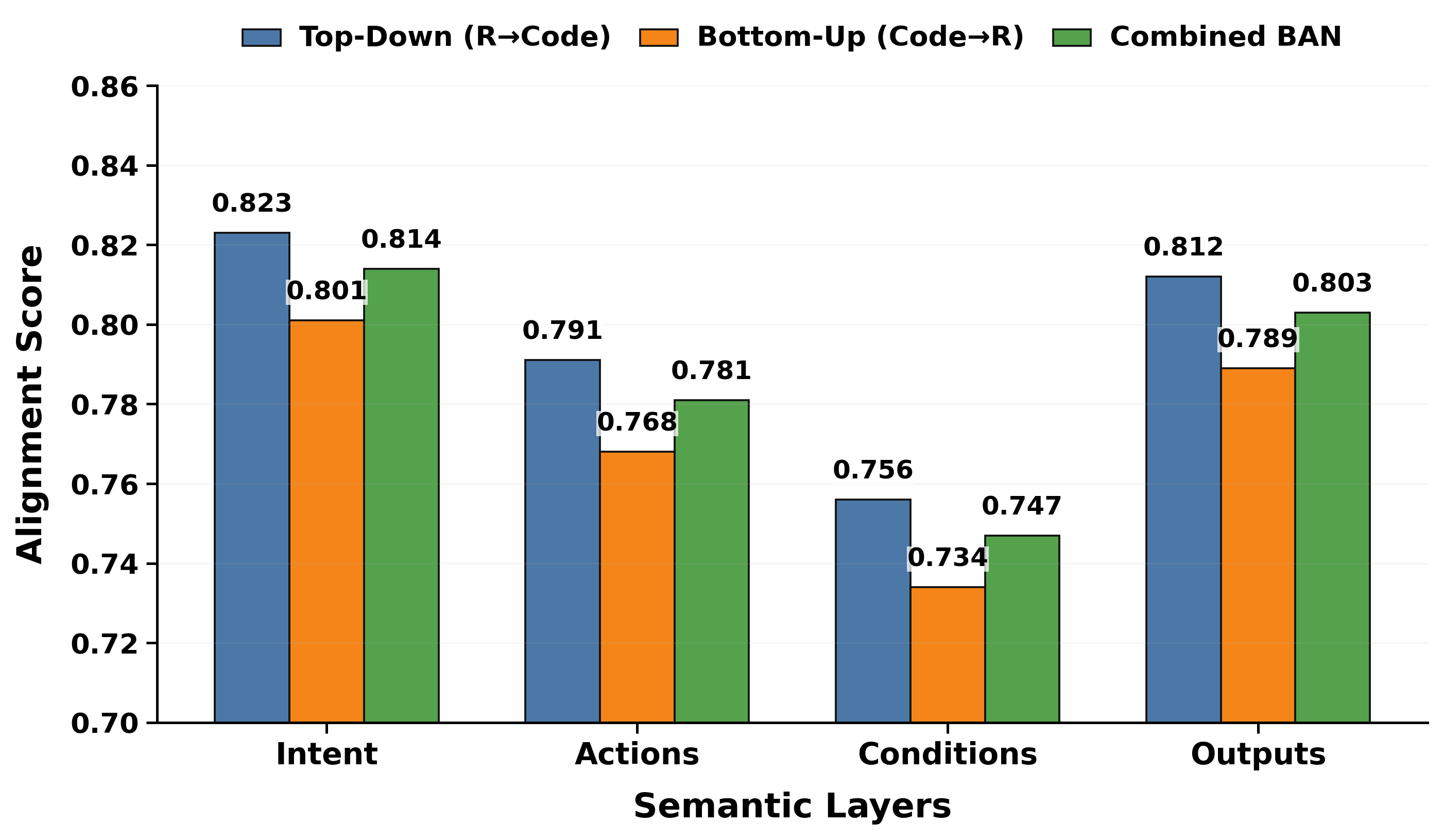}
    \caption{Layer-wise alignment scores of BAN on iTrust}
    \caption*{\footnotesize * Alignment score is the mean LLM-derived $sim(x,y)$ per layer pair over evaluated requirement--code pairs in iTrust.}
    \label{fig:hierarchical_layers}
\end{figure}

Fig.~\ref{fig:hierarchical_layers} reports BAN’s layer-wise alignment scores on iTrust across four semantic layers. 
The Intent layer achieves the highest alignment (0.814 in the combined setting), suggesting that high-level goal matching provides a strong traceability signal. The alignment decreases for more fine-grained layers, with Actions (0.781), Outputs (0.803), and Conditions (0.747), indicating increased semantic variability when matching detailed requirement semantics to code. 
In addition, the top-down (R$\rightarrow$Code) direction consistently yields higher alignment scores than bottom-up (Code$\rightarrow$R), while the combined strategy offers a more balanced signal across layers.

\begin{figure}[htbp]
    \centering
    \includegraphics[width=\linewidth]{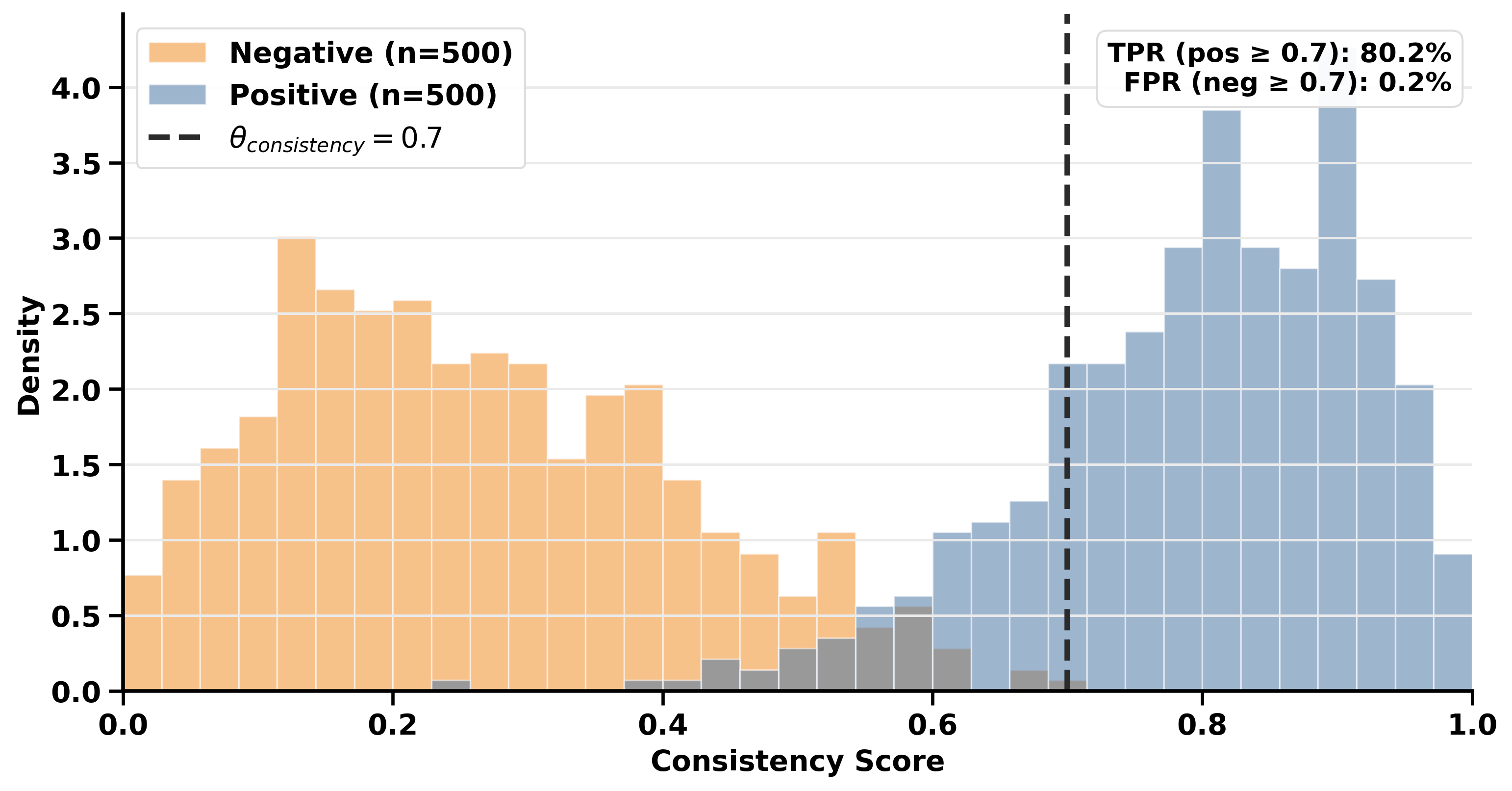}
    \caption{Discriminative consistency validation of SRCV}
    \caption*{\footnotesize * SRCV consistency score distributions for positive (ground-truth links) and negative (non-links) requirement--code pairs on iTrust ($n=500$ each). The dashed line indicates the filtering threshold $\theta_{\text{consistency}}=0.7$.}
    \label{fig:srcv_consistency}
\end{figure}

Fig.~\ref{fig:srcv_consistency} compares the SRCV consistency score distributions of positive and negative requirement--code pairs on iTrust. The positive pairs exhibit substantially higher consistency scores, while negative pairs are concentrated in the low-score region, indicating that SRCV provides a discriminative validation signal. Using the predefined threshold $\theta_{\text{consistency}}=0.7$, most negative candidates can be suppressed while preserving high-consistency positive links, supporting the effectiveness of SRCV for filtering spurious matches.

\subsection{Answer to RQ2: Robustness}
\begin{table}[htbp]
\caption{Cross-Dataset Performance Consistency}
\centering
\begin{tabular*}{\columnwidth}{l@{\extracolsep{\fill}}|c|c|c}
\toprule
Dataset & Best Baseline (F1) & R2Code (F1) & Gain \\
\midrule
iTrust & 0.6949 & 0.7296 & +5.0\% \\
eTour & 0.6812 & 0.7342 & +7.8\% \\
SMOS  & 0.6663 & 0.7201 & +8.1\% \\
eANCI & 0.7058 & 0.7592 & +7.6\% \\
RETRO & 0.6580 & 0.7134 & +8.4\% \\
\midrule
Average & 0.6812 & 0.7313 & +7.4\% \\
Std. Dev. & 0.0176 & 0.0157 & -- \\
\bottomrule
\end{tabular*}
\label{tab:cross_dataset}
\end{table}
To answer RQ2, Table~\ref{tab:cross_dataset}  reports the cross-dataset performance of R2Code under a fixed configuration, evaluating its robustness across five traceability datasets with diverse characteristics.
For each dataset, we compare R2Code against the strongest-performing baseline measured by F1-score. 
As shown in the table, R2Code consistently outperforms the best baseline on all datasets, achieving relative F1-score gains ranging from 5.0\% to 8.4\%. The improvements are stable across datasets, with an average gain of 7.4\% and comparable standard deviations between the baseline and R2Code results, indicating that the proposed approach maintains consistent effectiveness without dataset-specific tuning.

\subsection{Answer to RQ3: Efficiency}
% Analyze efficiency and cost trade-offs.
% Include Table 3 (Tokens / Time / Cost / F1) and Figure 3 (F1–Cost Curve).
% Discuss the balance between performance and computational cost.
\begin{table*}[t]
\centering
\caption{Efficiency and cost summary on the iTrust dataset}
\label{tab:eff_cost_summary}
\small
\setlength{\tabcolsep}{6pt}
\renewcommand{\arraystretch}{1.05}
\begin{tabular}{l c r r r r r r}
\toprule
Model & F1 &
\multicolumn{1}{c}{Input Tokens} &
\multicolumn{1}{c}{Output Tokens} &
\multicolumn{1}{c}{Runtime (s)} &
\multicolumn{1}{c}{Cost (USD)} &
\multicolumn{1}{c}{True Positives} &
\multicolumn{1}{c}{Cost/TP (USD)} \\
\midrule
BM25     & 0.6158 & 0         & 0       & 0.2     & 0.00   & 704   & 0.0000 \\
TF-IDF   & 0.6068 & 0         & 0       & 0.5     & 0.00   & 691   & 0.0000 \\
WMD      & 0.5910 & 0         & 0       & 1.8     & 0.00   & 680   & 0.0000 \\
RAG+LLM  & 0.6949 & 2,847,234 & 892,156 & 1,234.5 & 463.16 & 1,150 & 0.4027 \\
R2Code   & \textbf{0.7296} & 1,479,935 & 699,521 & 987.3   & \textbf{285.79} & 1,156 & \textbf{0.2472} \\
\bottomrule
\end{tabular}
\end{table*}

Table~\ref{tab:eff_cost_summary} summarizes the efficiency and cost results on the iTrust dataset. Traditional IR baselines (e.g., BM25 and TF-IDF) incur zero LLM API cost, but achieve substantially lower F1-scores. Compared with the RAG+LLM baseline, R2Code reduces total token consumption from 3.74M to 2.18M ($-41.7\%$), including a $48.0\%$ reduction in input tokens and a $21.6\%$ reduction in output tokens. This results in a lower estimated inference cost (\$463.16 to \$285.79, $-38.3\%$) and a shorter end-to-end runtime (1,234.5s to 987.3s, $-20.0\%$), while improving F1 from 0.6949 to 0.7296.

Table~\ref{tab:eff_cost_summary} further reports the cost per correct traceability link. R2Code achieves \$0.2472 per correct link with 1,156 true positives, whereas RAG+LLM requires \$0.4027 per correct link for 1,150 true positives. These results indicate that R2Code delivers higher traceability accuracy with lower per-link inference cost under the same evaluation setting.

\begin{table}[t]
\centering
\captionsetup{justification=raggedright,singlelinecheck=false}
\caption{Retrieval-stage efficiency: Fixed RAG vs. DCAR on iTrust} 
\label{tab:token_adapt_stats}
\begin{tabular}{l|c|c}
\toprule
\textbf{Metric} & \textbf{Fixed RAG} & \textbf{DCAR} \\
\midrule
Tokens per pair & 4,850 & 3,054 \\
Total tokens & 2,483,200 & 1,563,216 \\
\midrule
Avg. retrieved files (Low complexity) & -- & 8.3 \\
Avg. retrieved files (Medium complexity) & -- & 15.7 \\
Avg. retrieved files (High complexity) & -- & 23.4 \\
\bottomrule
\end{tabular}
\caption*{\footnotesize \textbf{Note.} Tokens are counted for retrieval construction only (excluding BAN/SRCV). “Retrieved files” denote pre-cap candidates (code entities).}
\end{table}
Table~\ref{tab:token_adapt_stats} summarizes token efficiency under fixed-window retrieval and DCAR on the iTrust dataset for the retrieval construction stage. 
DCAR reduces the average tokens per requirement--code pair from 4,850 to 3,054, corresponding to an approximate 37\% reduction. In terms of overall usage, total token consumption decreases from 2,483,200 to 1,563,216, indicating that adaptive context selection can substantially lower token overhead under the same evaluation setting.

Table~\ref{tab:token_adapt_stats} also reports how DCAR adjusts retrieval scope based on requirement complexity, measured as the number of retrieved candidate files.
Low-complexity requirements retrieve an average of 8.3 candidate files, medium complexity retrieves 15.7 candidate files, and high-complexity requirements expand to 23.4 candidate files.
This pattern shows that DCAR increases retrieval context for more complex requirements while keeping the number of files included in the final prompt bounded by \emph{kmax}=10.

%% file: 6_Discussion.tex
\section{Discussion}
% Summarize key findings and insights from experiments.
% Highlight implications and overall observations.

\subsection{Finding 1: Effectiveness}
R2Code’s effectiveness in RQ1 is best understood not as “more retrieval” but as better decision-making under noisy candidates: structured semantic alignment helps separate true implementation evidence from lexical look-alikes, making the top-ranked links more reliable for maintenance use. 
The ablation results reveal distinct and complementary roles. Removing BAN largely erases the advantage, indicating that cross-layer alignment is the primary source of improvement rather than incidental prompt effects.

Notably, removing SRCV can increase precision while substantially hurting recall. This suggests SRCV does more than filter false positives, it calibrates confidence so the system does not become overly conservative and miss valid links when evidence is partial or uneven. 
DCAR mainly improves evidence coverage. Removing it increases precision but reduces recall, indicating that dynamic evidence construction retrieves additional true links while also introducing harder, borderline candidates that BAN and SRCV help resolve.

\subsection{Finding 2: Robustness}
R2Code shows robust improvements across heterogeneous project structures and transfers to the C\# RETRO dataset under the same fixed configuration.
This robustness is consistent with decomposition-based alignment: by matching requirements and code at corresponding semantic layers, the framework is less dependent on shared vocabulary that may shift across domains or languages. 
In addition, SRCV contributes a domain-agnostic calibration signal. Fig.~\ref{fig:srcv_consistency}  shows that consistency scores separate positive from negative pairs and that a single threshold ($\theta=0.7$) suppresses most negative candidates while preserving high-consistency positives.
A remaining boundary case involves highly scattered implementations, where limited context budgets may miss intermediate evidence, suggesting more adaptive evidence expansion.

\subsection{Finding 3: Efficiency}
R2Code improves efficiency not simply by “using fewer tokens,” but by making the evidence budget more selective and task-aware. 
Compared with a fixed-window RAG+LLM pipeline, it reduces end-to-end runtime and estimated inference cost while still improving F1, indicating that traceability benefits more from compact, high-relevance evidence than from indiscriminately expanding context. 
Table~\ref{tab:token_adapt_stats} shows that DCAR expands retrieval scope in a structured way by allocating more candidate files to higher-complexity requirements (about 8.3 vs. 23.4 on average), while keeping the final prompt bounded by (\emph{kmax}=10). This suggests a practical scaling implication: efficiency gains come from removing irrelevant evidence and reserving broader context only for requirements that empirically need it.

\section{Threats to Validity}
\textbf{Internal validity.}
The reported results may be influenced by implementation and configuration choices, such as the selected LLM backbone and the prompting protocol, which can affect the precision--recall--cost trade-off in LLM-assisted pipelines.
To mitigate this threat, we fix the experimental setup across all datasets (same backbone, prompting protocol, and budgets) with deterministic decoding (e.g., temperature = 0) to reduce run-to-run variance, and we adopt a consistent setting based on a small pilot sensitivity analysis of prompt and configuration variants.
In addition, errors in LLM-generated intermediate representations may propagate to alignment and calibration, particularly for ambiguous requirements or highly dispersed implementations.

\textbf{External validity.}
The evaluation is conducted on public datasets with varying artifact granularity and link sparsity, which may influence absolute effectiveness. Cost estimates are derived from token usage and runtime measured in a specific environment, and may differ under alternative infrastructures or pricing schemes. In evolving repositories, cached code summaries may also become stale, motivating incremental update strategies for continuous integration settings.

%% file: 7_Conclusion.tex
\section{Conclusion}
In this paper, we proposed R2Code for requirements-to-code traceability, aiming to improve effectiveness, enhance efficiency, and maintain robust performance across heterogeneous projects. R2Code combines structured semantic decomposition, bidirectional alignment, confidence calibration, and context-adaptive evidence construction to produce reliable trace links under limited context budgets. Experiments on five public datasets (\textit{iTrust}, \textit{eTour}, \textit{SMOS}, \textit{eANCI}, and \textit{RETRO.NET}) show that R2Code consistently outperforms classical IR baselines and LLM-based retrieval pipelines, achieving an average relative F1 improvement of 7.4\% and up to 14.1\% on individual datasets. Meanwhile, R2Code reduces total token consumption by up to 41.7\%, and improves the cost per correct link by up to 37.0\%, demonstrating that better traceability quality can be obtained with substantially lower inference overhead.

Future work will extend R2Code to larger-scale industrial repositories and additional software artifacts, such as design documents, issue discussions, and commits, to further evaluate its applicability in more realistic development settings. We also plan to strengthen robustness in continuously evolving codebases by introducing incremental summary refresh and more adaptive evidence expansion for highly scattered implementations. In addition, we will explore more controllable decomposition and verification strategies to improve the reliability and interpretability of trace decisions. Finally, we aim to investigate multilingual traceability settings and broader cross-project scenarios to further enhance the generalization and practical adoption of the proposed framework.

%% file: 8_Ack.tex
\section*{Acknowledgment}
The work described in this paper is substantially supported by Hong Kong Metropolitan University Research Grant (Project Reference No. RD/2025/1.21 and No. RD/2025/1.24), and partially supported by the grant under the Research Grant Council (Project Reference No. UGC/FDS16/E25/25).

%% file: main.bib
@article{ramesh1997requirements,
  title={Requirements traceability: Theory and practice},
  author={Ramesh, Balasubramaniam and Stubbs, Curtis and Powers, Timothy and Edwards, Michael},
  journal={Annals of software engineering},
  volume={3},
  number={1},
  pages={397--415},
  year={1997},
  publisher={Springer}
}

@inproceedings{ghabi2012code,
  title={Code patterns for automatically validating requirements-to-code traces},
  author={Ghabi, Achraf and Egyed, Alexander},
  booktitle={Proceedings of the 27th IEEE/ACM International Conference on Automated Software Engineering},
  pages={200--209},
  year={2012}
}

@article{ramesh1998factors,
  title={Factors influencing requirements traceability practice},
  author={Ramesh, Balasubramaniam},
  journal={Communications of the ACM},
  volume={41},
  number={12},
  pages={37--44},
  year={1998},
  publisher={ACM New York, NY, USA}
}

@article{tian2021impact,
  title={The impact of traceability on software maintenance and evolution: A mapping study},
  author={Tian, Fangchao and Wang, Tianlu and Liang, Peng and Wang, Chong and Khan, Arif Ali and Babar, Muhammad Ali},
  journal={Journal of Software: Evolution and Process},
  volume={33},
  number={10},
  pages={e2374},
  year={2021},
  publisher={Wiley Online Library}
}

@article{lyu2023systematic,
  title={A systematic literature review of issue-based requirement traceability},
  author={Lyu, Yijing and Cho, Heetae and Jung, Pilsu and Lee, Seonah},
  journal={Ieee Access},
  volume={11},
  pages={13334--13348},
  year={2023},
  publisher={IEEE}
}

@inproceedings{hey2021improving,
  title={Improving traceability link recovery using fine-grained requirements-to-code relations},
  author={Hey, Tobias and Chen, Fei and Weigelt, Sebastian and Tichy, Walter F},
  booktitle={2021 IEEE International Conference on Software Maintenance and Evolution (ICSME)},
  pages={12--22},
  year={2021},
  organization={IEEE}
}

@incollection{guo2025natural,
  title={Natural language processing for requirements traceability},
  author={Guo, Jin LC and Stegh{\"o}fer, Jan-Philipp and Vogelsang, Andreas and Cleland-Huang, Jane},
  booktitle={Handbook on Natural Language Processing for Requirements Engineering},
  pages={89--116},
  year={2025},
  publisher={Springer}
}

@inproceedings{merten2016information,
  title={Do information retrieval algorithms for automated traceability perform effectively on issue tracking system data?},
  author={Merten, Thorsten and Kr{\"a}mer, Daniel and Mager, Bastian and Schell, Paul and B{\"u}rsner, Simone and Paech, Barbara},
  booktitle={International Working Conference on Requirements Engineering: Foundation for Software Quality},
  pages={45--62},
  year={2016},
  organization={Springer}
}

@inproceedings{north2024code,
  title={Code gradients: Towards automated traceability of llm-generated code},
  author={North, Marc and Atapour-Abarghouei, Amir and Bencomo, Nelly},
  booktitle={2024 IEEE 32nd International Requirements Engineering Conference (RE)},
  pages={321--329},
  year={2024},
  organization={IEEE}
}

@article{alor2025evaluating,
  title={Evaluating the Use of LLMs for Documentation to Code Traceability},
  author={Alor, Ebube and Khatoonabadi, SayedHassan and Shihab, Emad},
  journal={arXiv preprint arXiv:2506.16440},
  year={2025}
}

@article{zadenoori2025large,
  title={Large Language Models (LLMs) for Requirements Engineering (RE): A Systematic Literature Review},
  author={Zadenoori, Mohammad Amin and Dabrowski, Jacek and Alhoshan, Waad and Zhao, Liping and Ferrari, Alessio},
  journal={arXiv preprint arXiv:2509.11446},
  year={2025}
}

@article{wang2025hgnnlink,
  title={HGNNLink: recovering requirements-code traceability links with text and dependency-aware heterogeneous graph neural networks},
  author={Wang, Bangchao and Zou, Zhiyuan and Liang, Xuanxuan and Jin, Huan and Liang, Peng},
  journal={Automated Software Engineering},
  volume={32},
  number={2},
  pages={55},
  year={2025},
  publisher={Springer}
}

@inproceedings{marcus2003recovering,
  title={Recovering documentation-to-source-code traceability links using latent semantic indexing},
  author={Marcus, Andrian and Maletic, Jonathan I},
  booktitle={25th International Conference on Software Engineering, 2003. Proceedings.},
  pages={125--135},
  year={2003},
  organization={IEEE}
}

@article{sundaram2010assessing,
  title={Assessing traceability of software engineering artifacts},
  author={Sundaram, Senthil Karthikeyan and Hayes, Jane Huffman and Dekhtyar, Alex and Holbrook, E Ashlee},
  journal={Requirements engineering},
  volume={15},
  number={3},
  pages={313--335},
  year={2010},
  publisher={Springer}
}

@inproceedings{oliveto2010equivalence,
  title={On the equivalence of information retrieval methods for automated traceability link recovery},
  author={Oliveto, Rocco and Gethers, Malcom and Poshyvanyk, Denys and De Lucia, Andrea},
  booktitle={2010 IEEE 18th International Conference on Program Comprehension},
  pages={68--71},
  year={2010},
  organization={IEEE}
}

@inproceedings{guo2017semantically,
  title={Semantically enhanced software traceability using deep learning techniques},
  author={Guo, Jin and Cheng, Jinghui and Cleland-Huang, Jane},
  booktitle={2017 IEEE/ACM 39th International Conference on Software Engineering (ICSE)},
  pages={3--14},
  year={2017},
  organization={IEEE}
}

@inproceedings{lin2021traceability,
  title={Traceability transformed: Generating more accurate links with pre-trained bert models},
  author={Lin, Jinfeng and Liu, Yalin and Zeng, Qingkai and Jiang, Meng and Cleland-Huang, Jane},
  booktitle={2021 IEEE/ACM 43rd International Conference on Software Engineering (ICSE)},
  pages={324--335},
  year={2021},
  organization={IEEE}
}

@article{reimers2019sentence,
  title={Sentence-bert: Sentence embeddings using siamese bert-networks},
  author={Reimers, Nils and Gurevych, Iryna},
  journal={arXiv preprint arXiv:1908.10084},
  year={2019}
}

@article{feng2020codebert,
  title={Codebert: A pre-trained model for program-ming and natural languages},
  author={Feng, Z},
  journal={arXiv preprint arXiv:2002.08155},
  year={2020}
}

@article{guo2022unixcoder,
  title={Unixcoder: Unified cross-modal pre-training for code representation},
  author={Guo, Daya and Lu, Shuai and Duan, Nan and Wang, Yanlin and Zhou, Ming and Yin, Jian},
  journal={arXiv preprint arXiv:2203.03850},
  year={2022}
}

@inproceedings{fuchss2025lissa,
  title={LiSSA: toward generic traceability link recovery through retrieval-augmented generation},
  author={Fuch{\ss}, Dominik and Hey, Tobias and Keim, Jan and Liu, Haoyu and Ewald, Niklas and Thirolf, Tobias and Koziolek, Anne},
  booktitle={Proceedings of the IEEE/ACM 47th International Conference on Software Engineering. ICSE},
  volume={25},
  year={2025}
}

@inproceedings{hey2025requirements,
  title={Requirements Traceability Link Recovery via Retrieval-Augmented Generation},
  author={Hey, Tobias and Fuch{\ss}, Dominik and Keim, Jan and Koziolek, Anne},
  booktitle={International Working Conference on Requirements Engineering: Foundation for Software Quality},
  pages={381--397},
  year={2025},
  organization={Springer}
}

@inproceedings{ali2024establishing,
  title={Establishing traceability between natural language requirements and software artifacts by combining rag and llms},
  author={Ali, Syed Juned and Naganathan, Varun and Bork, Dominik},
  booktitle={International Conference on Conceptual Modeling},
  pages={295--314},
  year={2024},
  organization={Springer}
}

@article{leonhardt2024efficient,
  title={Efficient neural ranking using forward indexes and lightweight encoders},
  author={Leonhardt, Jurek and M{\"u}ller, Henrik and Rudra, Koustav and Khosla, Megha and Anand, Abhijit and Anand, Avishek},
  journal={ACM Transactions on Information Systems},
  volume={42},
  number={5},
  pages={1--34},
  year={2024},
  publisher={ACM New York, NY}
}

@article{tian2023cross,
  title={A cross-level requirement trace link update model based on bidirectional encoder representations from transformers},
  author={Tian, Jiahao and Zhang, Li and Lian, Xiaoli},
  journal={Mathematics},
  volume={11},
  number={3},
  pages={623},
  year={2023},
  publisher={MDPI}
}

@incollection{cleland2014software,
  title={Software traceability: trends and future directions},
  author={Cleland-Huang, Jane and Gotel, Orlena CZ and Huffman Hayes, Jane and M{\"a}der, Patrick and Zisman, Andrea},
  booktitle={Future of software engineering proceedings},
  pages={55--69},
  publisher = {ACM},
  year={2014}
}

@article{ramesh2002toward,
  title={Toward reference models for requirements traceability},
  author={Ramesh, Balasubramaniam and Jarke, Matthias},
  journal={IEEE transactions on software engineering},
  volume={27},
  number={1},
  pages={58--93},
  year={2002},
  publisher={IEEE}
}

@inproceedings{niu2016gray,
  title={Gray links in the use of requirements traceability},
  author={Niu, Nan and Wang, Wentao and Gupta, Arushi},
  booktitle={Proceedings of the 2016 24th ACM SIGSOFT international symposium on foundations of software engineering},
  pages={384--395},
  year={2016}
}

@inproceedings{rasiman2022effective,
  title={How effective is automated trace link recovery in model-driven development?},
  author={Rasiman, Randell and Dalpiaz, Fabiano and Espa{\~n}a, Sergio},
  booktitle={International Working Conference on Requirements Engineering: Foundation for Software Quality},
  pages={35--51},
  year={2022},
  organization={Springer}
}

@inproceedings{aung2020literature,
  title={A literature review of automatic traceability links recovery for software change impact analysis},
  author={Aung, Thazin Win Win and Huo, Huan and Sui, Yulei},
  booktitle={Proceedings of the 28th International Conference on Program Comprehension},
  pages={14--24},
  year={2020}
}

@inproceedings{zhang2016empirical,
  title={An empirical study on recovering requirement-to-code links},
  author={Zhang, Yuchen and Wan, Chengcheng and Jin, Bo},
  booktitle={2016 17th IEEE/ACIS International Conference on Software Engineering, Artificial Intelligence, Networking and Parallel/Distributed Computing (SNPD)},
  pages={121--126},
  year={2016},
  organization={IEEE}
}

@inproceedings{hey2024requirements,
  title={Requirements classification for traceability link recovery},
  author={Hey, Tobias and Keim, Jan and Corallo, Sophie},
  booktitle={2024 IEEE 32nd International Requirements Engineering Conference (RE)},
  pages={155--167},
  year={2024},
  organization={IEEE}
}

@inproceedings{wang2022systematic,
  title={A systematic mapping study of information retrieval approaches applied to requirements trace recovery.},
  author={Wang, Bangchao and Wang, Heng and Luo, Ruiqi and Zhang, Sen and Zhu, Qiang},
  booktitle={SEKE},
  pages={1--6},
  year={2022}
}

@inproceedings{hayes2003improving,
  title={Improving requirements tracing via information retrieval},
  author={Hayes, Jane Huffman and Dekhtyar, Alex and Osborne, James},
  booktitle={Proceedings. 11th IEEE International Requirements Engineering Conference, 2003.},
  pages={138--147},
  year={2003},
  organization={IEEE}
}

@inproceedings{zhang2023ealink,
  title={EALink: An efficient and accurate pre-trained framework for issue-commit link recovery},
  author={Zhang, Chenyuan and Wang, Yanlin and Wei, Zhao and Xu, Yong and Wang, Juhong and Li, Hui and Ji, Rongrong},
  booktitle={2023 38th IEEE/ACM International Conference on Automated Software Engineering (ASE)},
  pages={217--229},
  year={2023},
  organization={IEEE}
}

@article{borg2014recovering,
  title={Recovering from a decade: a systematic mapping of information retrieval approaches to software traceability},
  author={Borg, Markus and Runeson, Per and Ard{\"o}, Anders},
  journal={Empirical Software Engineering},
  volume={19},
  number={6},
  pages={1565--1616},
  year={2014},
  publisher={Springer}
}

@article{robertson2009probabilistic,
  title={The probabilistic relevance framework: BM25 and beyond},
  author={Robertson, Stephen and Zaragoza, Hugo and others},
  journal={Foundations and Trends{\textregistered} in Information Retrieval},
  volume={3},
  number={4},
  pages={333--389},
  year={2009},
  publisher={Now Publishers, Inc.}
}

@article{deerwester1990indexing,
  title={Indexing by latent semantic analysis},
  author={Deerwester, Scott and Dumais, Susan T and Furnas, George W and Landauer, Thomas K and Harshman, Richard},
  journal={Journal of the American society for information science},
  volume={41},
  number={6},
  pages={391--407},
  year={1990},
  publisher={Wiley Online Library}
}

@inproceedings{Kusner2015WMD,
  author    = {Kusner, Matt and Sun, Yu and Kolkin, Nicholas and Weinberger, Kilian},
  title     = {From Word Embeddings to Document Distances},
  booktitle = {Proceedings of the 32nd International Conference on Machine Learning (ICML)},
  year      = {2015},
  pages     = {957--966},
  publisher = {PMLR},
  url       = {https://proceedings.mlr.press/v37/kusnerb15.html}
}

@inproceedings{hayes2018requirements,
  title={The requirements tracing on target (retro). net dataset},
  author={Hayes, Jane Huffman and Dekhtyar, Alex and Payne, Jared},
  booktitle={2018 IEEE 26th International Requirements Engineering Conference (RE)},
  pages={424--427},
  year={2018},
  organization={IEEE}
}

@article{antoniol2025recovering,
  title={Recovering Traceability Links Between Code and Documentation: a Retrospective},
  author={Antoniol, Giuliano and Canfora, Gerardo and Casazza, Gerardo and De Lucia, Andrea and Merlo, Ettore},
  journal={IEEE Transactions on Software Engineering},
  year={2025},
  publisher={IEEE}
}

@article{lewis2020retrieval,
  title={Retrieval-augmented generation for knowledge-intensive nlp tasks},
  author={Lewis, Patrick and Perez, Ethan and Piktus, Aleksandra and Petroni, Fabio and Karpukhin, Vladimir and Goyal, Naman and K{\"u}ttler, Heinrich and Lewis, Mike and Yih, Wen-tau and Rockt{\"a}schel, Tim and others},
  journal={Advances in neural information processing systems},
  volume={33},
  pages={9459--9474},
  year={2020}
}

@misc{deepseek_json_output,
  author       = {{DeepSeek}},
  title        = {DeepSeek API Guide: JSON Output},
  howpublished = {\url{https://api-docs.deepseek.com/guides/json_mode}},
  note         = {Accessed: 2026-02-05}
}

@article{li2024simac,
  title={SimAC: simulating agile collaboration to generate acceptance criteria in user story elaboration},
  author={Li, Yishu and Keung, Jacky and Yang, Zhen and Ma, Xiaoxue and Zhang, Jingyu and Liu, Shuo},
  journal={Automated Software Engineering},
  volume={31},
  number={2},
  pages={55},
  year={2024},
  publisher={Springer}
}

@inproceedings{li2024llm,
  title={Llm-based class diagram derivation from user stories with chain-of-thought promptings},
  author={Li, Yishu and Keung, Jacky and Ma, Xiaoxue and Chong, Chun Yong and Zhang, Jingyu and Liao, Yihan},
  booktitle={2024 IEEE 48th Annual Computers, Software, and Applications Conference (COMPSAC)},
  pages={45--50},
  year={2024},
  organization={IEEE}
}

@article{wang2025chart2code,
  title={Chart2Code-MoLA: Efficient Multi-Modal Code Generation via Adaptive Expert Routing},
  author={Wang, Yifei and Keung, Jacky and Mao, Zhenyu and Zhang, Jingyu and Cao, Yuchen},
  journal={arXiv preprint arXiv:2511.23321},
  year={2025}
}

@article{mao2025re,
  title={Towards Requirements Engineering for GenAI-Enabled Software: Bridging Responsibility Gaps through Human Oversight Requirements},
  author={Mao, Zhenyu and Keung, Jacky and Sun, Yicheng and Wang, Yifei and Liu, Shuo and Li, Jialong},
  journal={arXiv preprint arXiv:2511.13069},
  year={2025}
}

@article{mao2025multiagent,
  title={Towards Engineering Multi-Agent LLMs: A Protocol-Driven Approach},
  author={Mao, Zhenyu and Keung, Jacky and Zhang, Fengji and Liu, Shuo and Wang, Yifei and Li, Jialong},
  journal={arXiv preprint arXiv:2510.12120},
  year={2025}
}
